\documentclass{appolb}
\usepackage{epsfig}

\begin{document}
\title{Diffractive Vector Meson Production in
 $k_t$-Factorization Approach
\thanks{Presented at X International Workshop on Deep Inelastic Scattering (DIS2002)
Cracow, Poland, 30 April - 4 May 2002}%
}
\author{I.P.~Ivanov
\address{Forschungszentrum Juelich, Germany and Novosibirsk University, Russia}
\and
N.N.~Nikolaev
\address{Forschungszentrum Juelich, Germany and ITP, Moscow}
}
\maketitle
\begin{abstract}
We describe the current status of the diffractive vector meson
production calculations within the $k_t$ factorization approach.
Since the amplitude of the vector meson production off a proton is
expressed via the differential gluon structure function (DGSF), we
take a closer look at the latter and present results of our new
improved determination of the DGSF from the structure function
$F_{2p}$. Having determined the differential glue, we proceed to
the $k_t$-factorization results for the production of various
vector mesons. We argue that the properties of the vector meson
production can reveal the internal spin-angular and radial
structure of the vector meson.
\end{abstract} 
\section{Introduction}

Diffractive DIS appears to be a unique probe of the internal
structure of hadrons in the Regge regime, \ie at $Q^2$ fixed, $x
\to 0$. It has been always appreciated that diffractive processes
can give valuable information of the gluon content inside a
rapidly moving proton. Another aspect  might not be that obvious,
namely, that diffraction can tell us much about the structure of
the produced system as well.

The case of the vector meson production $$ \gamma^{(*)}p \to Vp $$
is especially promising.
A general expression for transition $\gamma^{(*)}(\lambda_\gamma) \to
V(\lambda_V)$ is given by:
\begin{eqnarray}
{\cal A}(x,Q^{2},\vec \Delta)= is{c_{V}\sqrt{4\pi\alpha_{em}}
\over 4\pi^{2}} \int_{0}^{1} {dz\over z(1-z)} \int d^2 \vec k
\psi(z,\vec k) \nonumber\\ \cdot\int {d^{2} \vec \kappa \over
\vec\kappa^{4}}\alpha_{S}(q^2) \left(1 + i {\pi \over 2}
{\partial\over \partial\log x}\right) {\cal{F}}(x,\xi,\vec
\kappa,\vec \Delta) \cdot I(\gamma^{*}\to V)\nonumber
\end{eqnarray}
Here we have two quantities that are not calculable within pQCD,
namely, the radial wave function of the vector meson $\psi(z, \vec
k)$ and the off-forward unintegrated gluon density
${\cal{F}}(x,\xi,\vec \kappa,\vec \Delta)$.

By studying the properties of the
production of vector mesons with helicity $\lambda_V$ from photons
with helicity $\lambda_\gamma$, one can peep into the spin-angular
structure of the quarks inside the vector meson. Indeed, even the
mere fact that helicity-flip $\gamma \to V$ transitions take place
implies a certain motion of the constituents inside the meson.

It turns out that the distribution of the absolute values of the
relative momenta between the quark and the antiquark inside the
vector meson, the quantity that we just called the wave function of
the vector meson, does leave signatures in the properties of the
diffractive production as well. In our work we do not aim at the
most accurate possible way of describing the vector meson wave
function. What we do is we take two radically different Ans\"atze
--- the Coulomb and the oscillator type --- for the wave function
and check how different the predictions based on the form of the
wave function are. This will give us a reliable estimate of the
sensitivity of the whole approach to the details of the wave
function.

\section{Unintegrated gluon density}

Although the unintegrated gluon structure function is not
calculable from first principles, it can be extracted from
experimental data on $F_{2p}$. This analysis was performed in
\cite{IN1} and yielded compact and ready-to-use parametrizations
of ${\cal F}(x,\vec\kappa^2)$. Since this analysis was based on
simple eye-ball fits and because new data appeared during last two
years, we re-extracted the unintegrated gluon density from
$F_{2p}$, this time by means of $\chi^2$-minimization. We used 191
data points in low to moderate $Q^2$ ($Q^2 < 10$ GeV$^2$),
$x<0.01$ region and obtained three fits with
$\chi^2/n_{d.o.f.}\approx 1.25\div 1.40$. We hope that later on,
when calculating vector meson production cross sections, using
these three fits will give us a reliable estimate of the level of
uncertainty introduced by unintegrated glue.

This minimization procedure resulted in somewhat unexpected
two-peak shape of the unintegrated glue, which means that the soft
and hard components of the pomeron are strongly separated. Further
analysis is needed in order to establish whether the data indeed
favor the non-monotonous $\vec\kappa^2$-behavior of ${\cal
F}(x,\vec\kappa^2)$ or this is an artefact.

\section{Production of ground state mesons}

The $k_t$-factorization calculations of the vector meson
production cross section are known to possess a {\em scaling
phenomenon}: all ground state vector mesons, when corrected to
flavor factor, follow the same $Q^2+m_V^2$ dependence. Moreover,
we predict that a similar phenomenon should take place also for
the energy rise exponent and for the diffractive slope.

\begin{figure}[!htb]
   \centering
   \epsfig{file=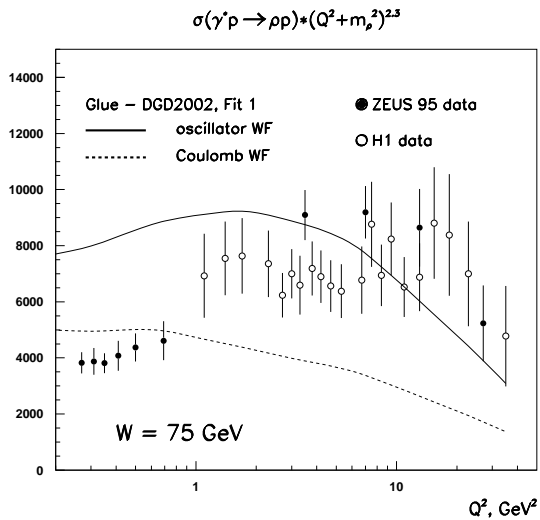,width=8cm}
   \caption{The $\rho$ meson production cross section multiplied
   by $(Q^2+m_\rho^2)^{2.3}$. Discrepancies at $Q^2 \sim 1$ GeV$^2$
   and at $Q^2 > 10$ GeV$^2$ can be clearly seen.}
\end{figure}

When calculating the production of $\rho$-meson and comparing them
with the data, we observed a good overall agreement except for two
$Q^2$ regions. Fig.1 display the quantity $\sigma(\gamma^*p\to\rho
p)\cdot(Q^2+m_\rho^2)^{2.3}$. At low $Q^2$ we observe the
experimental data points drop sharply, while our predictions do
not exhibit this behavior. It must be emphasized that this
discrepancy is located at $Q^2 \sim 1$ GeV$^2$, {\em not at  very
small $Q^2$}. In fact, with oscillator wave function, we managed
to reproduce the $Q^2 \to 0$ behavior of the total cross section
nearly perfectly.

At large enough $Q^2$ we observe another deviation between
$k_t$-factorization prediction and the data. As seen in Fig.1, at
high $Q^2$ both our predictions for  $\sigma(\gamma^*p\to\rho
p)\cdot(Q^2+m_\rho^2)^{2.3}$ and the data decrease, but our curves
start decreasing {\em earlier}. This problem was dubbed by us the
$\sigma_T$ puzzle, since it becomes even more glaring when
$\sigma_T$ and $\sigma_L$ are analyzed separately. During the last
two years, a number of ideas was investigated --- namely, the
effect of hard Coulomb tail of the wave function and strong
$S$-wave/$D$-wave mixing, see \cite{IN2} --- which could in
principle cure this problem. However, no satisfactory solution was
found.

If we ask not for the absolute values of cross section, but for
relative strength of various helicity amplitudes, the agreement
between $k_t$-factorization prediction and the data is much
better. For example, our calculation of the $\rho$-meson spin
density matrix and of the ratios spin-flip/non-spin-flip yielded
numbers rather close to data points. This make us believe that we
grasped the essential part of the spin physics of the diffractive
$\gamma\to\rho$ transition.

\begin{figure}[!htb]
   \centering
   \epsfig{file=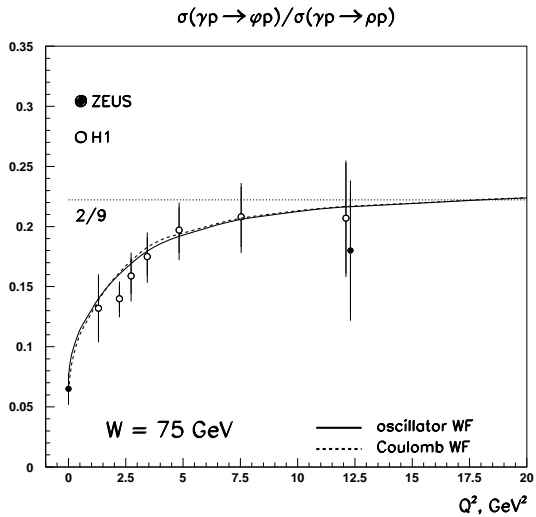,width=8cm}
   \caption{The ratio $\sigma(\phi)/\sigma(\rho)$ as function
   of $Q^2$.}
\end{figure}
In the case of other ground state vector mesons ($\omega, \phi,
J/\psi$), the $k_t$-factorization predictions were in a good
overall agreement with the data. Here we wish to show one
particularly interesting quantity, namely, the ratio of
$\phi$-meson to $\rho$-meson cross sections, taken at equal $Q^2$,
Fig.2. Aside from the observation that our predictions are in a
nice agreement with the data, we notice that predictions based on
Coulomb and oscillator wave functions virtually coincide. This is
highly non-trivial, since these two Ans\"atze lead to {\em very
different} absolute values of the cross section. This observation
suggests that quantity $\sigma(\phi)/\sigma(\rho)$ is practically
insensitive to the details of the internal structure of the vector
mesons (provided it is the same for $\rho$ and $\phi$). Therefore,
in some sense, this quantity appears to be a {\em parameter-free}
prediction of the theory, and the fact that it agrees with the
data supports our approach. An interesting issue remains to be
settled, namely, what is this quantity mostly sensitive to.

\section{Production of excited vector mesons}

Since we explicitly take into account the spin-angular coupling
inside the vector meson, we can calculate production of pure
$S$-wave or pure $D$-wave vector mesons, as well as any given
$S/D$ wave mixture. During such calculations, we observed a number
of remarkable phenomena. In the case of $2S$ states (radial
excitations) we observed the famous {\em node effect} and studied
its influence on production cross sections. We observed
characteristic $Q^2$ and $t$ shape of the (differential) cross
sections. Whenever possible, we compared predictions with the data
and found good agreement. In the case of $D$ wave state we found
strong overall suppression that comes from orthogonality between
the pure $D$ wave in vector meson and the dominant $S$ wave in the
initial photon.

We also calculated the spin density matrices and
$\sigma_L/\sigma_T$ ratios for all these states. The most
spectacular quantity here is $\sigma_L/\sigma_T$. For $2S$ state
in $\rho$ system, we observed a strongly oscillating $Q^2$ shape
of this ratio in the small $Q^2$ region, which is again
manifestation of the node effect. For $D$ wave state we found an
overall suppression of this ration in comparison with $1S$ state
by more than an order of magnitude.

We think that study of excited states, especially in the $\rho$
system, will shed light on the nature and internal structure of
various $\rho'$ resonances. In this aspect, HERA will yield
information hadron structure, complementary to studied at $e^+e^-$
colliders.

\end{document}